\documentclass{article}
\usepackage{spconf}
\usepackage[bottom]{footmisc}
\usepackage{graphicx}
\usepackage{amssymb,amsmath,bm}
\usepackage{textcomp}
\usepackage{graphicx}
\usepackage{wrapfig}
\usepackage{lipsum}
\usepackage{dblfloatfix}
\usepackage{fixltx2e}
\usepackage{nameref}
\usepackage{graphicx}
\usepackage{algorithm}
\usepackage{varioref}
\usepackage{booktabs} 
\usepackage[noend]{algpseudocode}
\usepackage{numprint}
\usepackage{textcomp}
\usepackage[T1]{fontenc}
\def\BState{\State\hskip-\ALG@thistlm}

\usepackage{algpseudocode}
\usepackage[noend]{algpseudocode}
\usepackage{tablefootnote}

\usepackage{xcolor}

\title{CURE Dataset: Ladder Networks for Audio Event Classification}
\name{Harishchandra Dubey$^1$\textsuperscript{+}~\thanks{\textcolor{blue}{This material is presented to ensure timely dissemination of scholarly and technical work. Copyright and all rights therein are retained by the authors or by the respective copyright holders. The original citation of this paper is: H. Dubey, D. Emmanouilidou and I. J. Tashev , "CURE Dataset: Ladder Networks for Audio Event Classification", IEEE Pacific Rim Conference on Communications, Computers and Signal Processing (PacRim), August 21-23, 2019, Victoria, Canada. }}~\thanks{\textsuperscript{+}The work was done while H. Dubey was an intern at Microsoft Research Labs, Redmond, USA and PhD student at the University of Texas at Dallas, USA. H. Dubey is currently with Microsoft Corporation, Redmond, USA.}, Dimitra Emmanouilidou$^2$, Ivan J. Tashev$^2$}
\address{$^1$Microsoft Corporation, Redmond, WA, USA\\
	$^2$Microsoft Research, Redmond, WA, USA \\
	{\small \tt \{hadubey, diemmano, ivantash\}@microsoft.com}
}
\begin{document}
\maketitle
\begin{abstract}
%
Audio event classification is an important task for several applications such as surveillance, audio, video and multimedia retrieval etc. There are approximately 3M people with hearing loss who can't perceive events happening around them. This paper establishes the CURE dataset which contains curated set of specific audio events most relevant for people with hearing loss. We propose a ladder network based audio event classifier that utilizes 5s sound recordings derived from the Freesound project. We adopted the state-of-the-art convolutional neural network (CNN) embeddings as audio features for this task. We also investigate extreme learning machine (ELM) for event classification. In this study, proposed classifiers are compared with support vector machine (SVM) baseline. We propose signal and feature normalization that aims to reduce the mismatch between different recordings scenarios.  
Firstly, CNN is trained on weakly labeled Audioset data. Next, the pre-trained model is adopted as feature extractor for proposed CURE corpus. We incorporate ESC-50 dataset as second evaluation set. 
Results and discussions validate the superiority of Ladder network over ELM and SVM classifier in terms of robustness and increased classification accuracy. While Ladder network is robust to data mismatches, simpler SVM and ELM classifiers are sensitive to such mismatches, where the proposed normalization techniques can play an important role. Experimental studies with ESC-50 and CURE corpora elucidate the differences in dataset complexity and robustness offered by proposed approaches.
\end{abstract}
%
%
%
%
%
%
\section{Introduction}
\label{sec:intro}
Audio event  classification is an emerging research topic with applications in ambient sensing, low-hearing support systems, surveillance and home security monitoring, multimedia information retrieval, and many others~\cite{atrey2006audio}.
Detecting sound events in daily-life scenarios can be a challenging task: (a) in nature, sound events of interest can exhibit similar temporal and spectral characteristics and can overlap both in time and content; (b) presence of external or inherent interference, including ambient noise, recording noise and room reverberation, can suppress discriminatory information or alter the acoustic signature of underlying events. These challenges can lead to inevitable mismatch in training and test conditions.
In recent years, deep learning advancements, publicly available datasets~\cite{gemmeke2017audio, piczak2015environmental, salamon2014dataset, mesaros2016tut}, as well as scientific competitions~\cite{mesaros2018classification,kong2018dcase} contributed to rapid developments in the field.

In recent literature, conventional feature extraction approaches, such as Mel-frequency cepstral coefficients (MFCC), even when combined with ensemble classification models, displayed lower discriminatory capabilities than expected; thus exposing the need for more sophisticated features for audio event classification tasks~\cite{piczak2015environmental}.
Several deep neural network (DNN) based approaches have been proposed for feature extraction, among which convolutional neural networks (CNN) have shown increased performance~\cite{kumar2016audio, cakir2017convolutional}. CNNs provide robust data-driven feature representations that can potentially alleviate the need for excessively complex classification models. Kumar et al. demonstrated this idea in~\cite{kumar2017knowledge} with a deep CNN network originally designed for large-scale image recognition tasks, and expanded with additional layers for the task at hand. Similar observations were made by Aytar et al.~\cite{aytar2016soundnet}, and Piczak~\cite{piczak2015environmental} where CNNs achieved superior performance than feed-forward DNN counterparts and were able to improve classification accuracy from 73\% to 83\%. Recent work in ~\cite{jung2017dnn}, as part of DCASE challenge, demonstrated the complications of over-fitting when using a fusion of complex feature spaces with deep neural architectures.
%
\begin{figure*}[!t]
\centering
\includegraphics[width=1.0\textwidth, trim={30mm 0mm 0mm 0mm},clip]{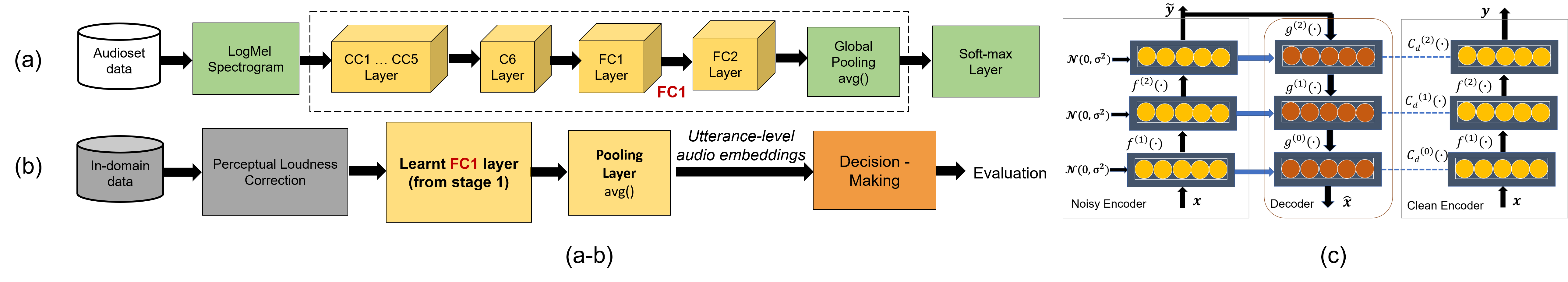}
\caption{(a-b) Proposed framework: Top sub-figure in (a-b) depicts the deep CNN architecture in the vanilla transfer learning pipeline. Audio embeddings were extracted from layer FC1 of the network, trained on  out-of-domain data. Layers CC1...CC5 are double convolutional layers and C6 is a single convolutional layer. FC1 layer includes batch normalization and ReLu activation while FC2 has sigmoid activation. The global pooling layer averages segment-level embeddings to utterance-level. The soft-max layer outputs predicts the posteriors for each audio event class. Bottom sub-figure in (a-b) illustrates the decision-making pipeline that leverages audio embedding (FC1) extracted from entire audio clips in evaluation set. (c) Illustration of a two-layer LadderNet, where $\mathbf{x}$ and $\widehat{\mathbf{x}}$ are input and reconstructed embeddings, $\mathbf{y}$ is the output label, and $\widetilde{\mathbf{y}}$ the output of noisy encoder, injected by Gaussian noise $\mathcal{N}(0,\sigma^{2})$. Decoder paths are characterized by denoising functions $\mathbf{g^{(.)}}$ and denoising costs $\mathbf{C_{d}^{(.)}}$ at each layer.}
\label{fig_proposed}
\vspace{-3mm}
\end{figure*}
This paper presents an audio event classification scheme by exploring architectures that do not assume uniformity between training and test data, and accounts for realistic constraints in model complexity and expected system latency. The contributions of this paper are as follows: (i) establish a new data set that contains a balanced set of audio events, relevant for real-world applications, obtained under varied non-uniform acoustic conditions, and labelled having low-latency applications in mind; (ii) explore vanilla transfer learning (VTL) for extracting efficient audio embeddings from out of domain data without the need for model re-training and additional layer adjustments for in-domain data;
(iii) investigate normalization procedures and propose best practices for tackling data mismatch and achieving increased model-specific performance;
 (iv) explore Ladder Network (LadderNet) and Extreme Learning Machine (ELM) for audio event classification and propose best system for specific infrastructure needs.
Section~\ref{sec:data} summarizes the datasets used in this study. Proposed approaches are described in Section~\ref{sec:methods} and experimental results presented in Section~\ref{sec:exp}. The paper concludes with a summary and discussion in Section~\ref{sec:discuss}.
%
\section{Data}
\label{sec:data}
\subsection{AudioSet}
\label{sec:outdomain}
%
AudioSet is a large-scale weakly labeled collection of $10$~sec long audio recordings extracted from about $2$~million YouTube videos, and grouped into $527$ highly unbalanced classes of at least $59$ recordings per class~\cite{gemmeke2017audio}. A loose hierarchy was used to categorize the sounds, and audio clips were weakly labeled containing one or more audio events with unknown start and end time. Data were organized into fixed training and validation sets by the authors and a balanced set was preferred for this work. Since many annotators label each clip, usually it is imperative to estimate inter-annotator agreement such as the correlation (or some other measures of agreement) between labels done by different annotators for same clip. AudioSet is a weakly labeled database where inter-annotator correlation was not computed. While there were no provided labels with cross-annotator agreement guarantees (a tedious process for large corpora), subsets of recordings underwent quality assessment by randomly sampling all thematically distinct classes. Depending on the application, the duration of these recordings can be too long to use for audio event classification systems that are required to respond in real-time situations. This corpus was used for  out-of-domain feature learning.
%
\subsection{Environmental Sound Classification (ESC-50)}
\label{sec:indomain}
ESC-50 is a labeled collection of 2000 public field audio recordings with $5$~sec duration, split into $50$ semantical classes under the theme of animals, natural soundscapes, human non-speech sounds, domestic sounds and exterior noises~\cite{piczak2015esc}. All classes are balanced containing $40$ examples each. A crowd sourcing platform was used for data annotation presenting reviewers with $50$ choices to choose from. ESC-50 was used in this work as part of in-domain data, with fixed train and test sets provided by the authors.
%
\subsection{Proposed CURE Corpus}
\vspace{-3mm}
We established a corpus containing \textit{CUR}ated set of specific audio \textit{E}vents (CURE) in which the audio clips were selected from a specific set of audio events. These audio events were chosen to capture most relevant events for people with hearing loss and consists of 13 classes. Unlike Audioset and ESC-50 dataset, CURE corpus consist of 5s audio recordings that are annotated by human transcribers to generate the ground-truth audio event labels. In this way, it facilitate study of audio events classification using short 5s clip that is not possible with Audioset and ESC-50. While ESC-50 contains 10s audio recordings for single event, Audioset has variable duration with one or more events. Thus, CURE corpus is complementary to existing audio event datasets. Datasets used in DCASE challenges~\cite{mesaros2017dcase} target different domains excluding people with hearing loss. CURE corpus is developed for research audio event classifier for 5s recordings on smartphone. There are 3 Millions individuals with hearing loss and that shows the scope of CURE corpus. This paper describes the data collection and annotation along with novel Ladder Network classifier for robust event classification. Implementation on smartphone is beyond the scope of this paper. Contrary to other available datasets, CURE corpus is designed to contain practically relevant sound events, chosen for their increased sensory effect: their ability to spur strong stimulation effects on users. Audio clips were queried from Freesound online repository, noted for its diversity in data capturing modalities, and variety in acoustical room conditions and environments~\cite{font2013freesound}. Queried sounds were segmented into $5$~sec clips without overlap, creating a pool of more than $7,000$ audio clips. The choice of $5$~sec duration is important and is meant to allow for enough context without blending out audio signatures of various events. It also enables system designing with realistic latency requirements, hence making it a reasonable choice. Audio clips were annotated by three judges using UHRS crowd sourcing~\cite{UHRS} into one of $13$ classes: fire and smoke alarm, baby crying and baby vocalizations, car horn, hand clapping, clock-alarm, human coughing, dog barking and dog howling, door bell ringing, door knocking, foot steps, glass breaking, ambulance or police siren, and thunderstorm sounds.
Judges were restricted to $100$ files to avoid bias and ground-truth was obtained when at least two judges agreed on the label. A balanced final set was achieved containing $500$ sounds per class. Human accuracy was calculated by accounting for the following cases as errors: when only two judges agreed on a label (counts as $1$ error) and when all judges disagreed (counts as $3$ errors). CURE dataset is target domain in this paper and hence used for evaluation experiments.
%
\section{Methods}
\label{sec:methods}
\vspace{-2mm}
\subsection{Data Preparation}
\label{sec:Data}
\vspace{-2mm}
Data were re-sampled to $16$~kHz, and further treatment was needed for the two in-domain sets.
Sound data coming from a variety of sources typically has variability in recording equipment, acoustical environment and noise levels. When designing frameworks that can perform and adapt to realistic applications, it is important to avoid highly uneven conditions; this is also important for human annotators, where acoustic variations may incur unwanted bias and skew the perceived contextual information. Since this can highly affect the new dataset, CURE was pre-treated with perceptual and loudness correction using offset correction and frame-level A-weighted filtering before annotation and usage.


We further propose a feature-level normalization scheme, useful for efficient learning and classification in a variety of settings. A whitening process is applied to both training and test feature data using the training set mean and variance. Whitening was followed by length-normalization $\hat{\mathbf{x}} = \frac{\mathbf{x}}{\|\mathbf{x}\|}$ that aims towards a compact feature representation lying on a unit hypersphere, where $\hat{\mathbf{x}}$ is the normalized feature vector of ${\mathbf{x}}$, and $\|.\|$ represents the vector norm. Combined steps of whitening and length-normalization are simply referred to as normalization, or feature-normalization in this manuscript.
%
%
\vspace{-2mm}
\subsection{Feature Extraction}
%
Audio event detection can be done on few corpora as discussed in earlier section. For research discussed in this paper, CURE corpus is our target application i.e., in-domain data. We leveraged AudioSet data for transfer learning based on Convolutional Neural Network (CNN). Thus, AudioSet is our out-of-domain data as it contains audio extracted from web videos. ESC-50 dataset is used as another evaluation set in addition to CURE corpus. ESC-50 is a out-of-domain evaluation set for research discussed in this paper.  
Transfer learning refers to the re-usage of knowledge coming from an  out-of-domain task to an in-domain related setting. It is an indirect way of reusing the plurality of data and information that a related domain can offer. A deep CNN architecture inspired by the computer vision literature was adopted for vanilla transfer learning (VTL)~\cite{kumar2017knowledge, oquab2014learning}. The network was trained on the large, balanced,  out-of-domain corpus AudioSet (section \ref{sec:outdomain}). Log Mel-spectral features were extracted over $1.5$~sec frames with $50$\% overlap to train the feature extracting network (Fig.~\ref{fig_proposed}a). Segment-level features were pooled to get utterance-level features before the soft-max output layer of dimension $527$ (number of  out-of-domain classes). Multi-label training loss was computed by averaging the binary cross-entropy loss per class. Network parameters were optimized using the validation set, and the resulting network was adopted for feature extraction at the FC1 layer (Fig.~\ref{fig_proposed}a), without the need for re-training or further adjustment to in-domain data. Extracting learned features from layer FC1 (VTL-CNN) is a great enabler for processing variable-length real-time streams of audio, by providing fixed-length embeddings.
The final segment-level features were extracted using the in-domain data (CURE) or ESC-50 at layer FC1, and then pooled to utterance-level features for usage by the decision-making models.
%
%
%
\subsection{Decision Making}
%
The audio event classification task is seen here as a classification problem.
The proposed models were used in conjunction with the accompanied in-domain fixed-length data sets, but are straight-forward to be generalized as pseudo real-time predictors for continuous audio stream processing.
%
\subsubsection{Ladder Network}
%
Ladder network (LadderNet) is a semi-supervised model designed to learn efficiently from small amounts of annotated data and has the ability to extrapolate information from unlabeled data with the assumption of low intra-class distance among feature samples~\cite{rasmus2015semi}. It comprises of three paths: clean encoder, noisy encoder, and decoder (Fig.~\ref{fig_proposed}c). Training and evaluation of the network is performed using the noisy encoder, which regularizes the input by introducing zero-mean Gaussian noise $\mathcal{N}(0,\sigma^2)$ to the noisy hidden layers and decoder path. Each layer of the decoder combines two signals, one from the layer above, and one from the corresponding layer in the noisy encoder. Lateral skip connections, a notion equivalent to hierarchical latent variable modeling~\cite{pezeshki2016deconstructing}, allow for sharing representation between the noisy encoder and corresponding decoder layers. The final prediction and testing is achieved via the clean encoder path. The lower layers of the network focus on accurate reconstruction between input $x$ and reconstruction $\hat{x}$, while the top layers learn to predict the output label from the given input features.
\begin{figure*}[!h]
\centering
\includegraphics[width=1.0\textwidth, trim={2mm 6mm 4mm 6mm},clip]{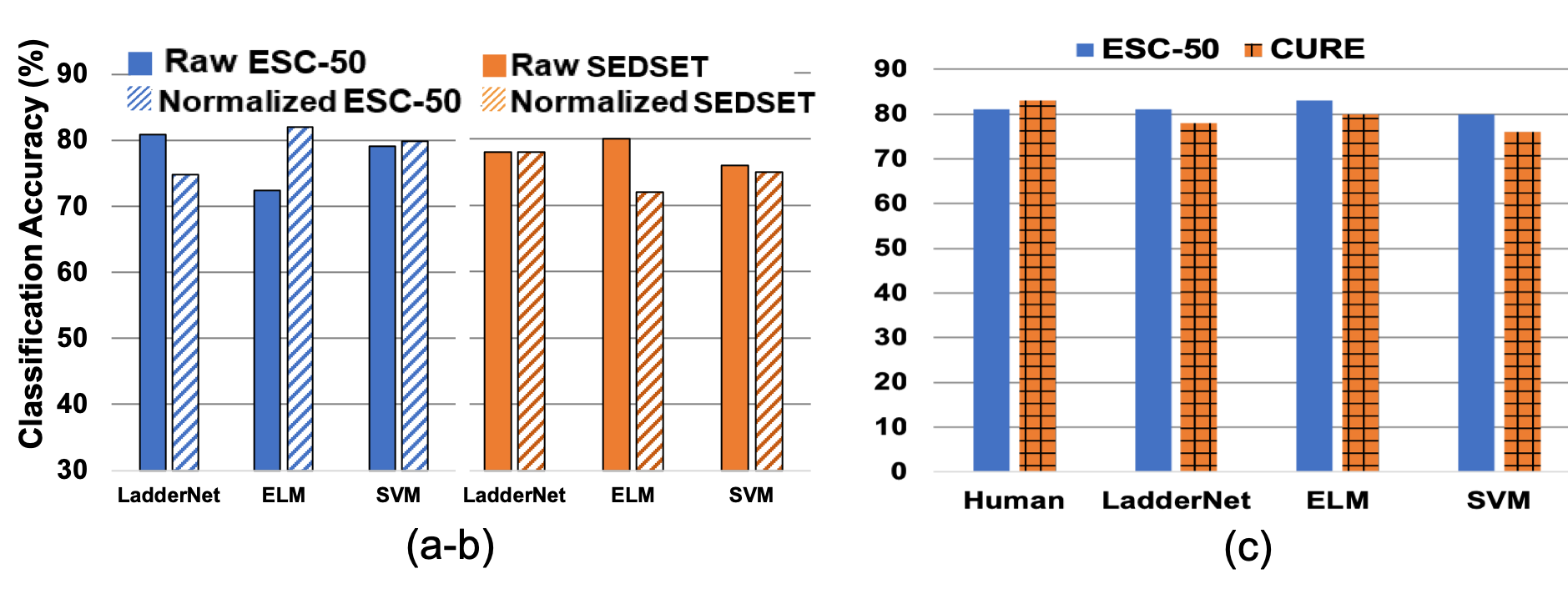}
\caption{(a-b) Effect of feature-based normalization on weighted classification accuracy (\%) on ESC-50 and CURE dataset; Raw represents the case where audio embeddings were used as it is. Normalized refers case where audio embeddings were length-normalized leading to unit-length feature vectors. Prior to length-normalization, we also performed mean and variance normalization; (c) Best classification accuracy of proposed models for two evaluation dataset.}
\label{fig_results_all}
\end{figure*}
%
\subsubsection{Extreme Learning Machine}
Extreme Learning Machine (ELM) is a single hidden layer neural network meant for simple and efficient learning~\cite{huang2006extreme}. Unlike traditional neural networks, ELMs can work with non-differentiable activation functions and have random hidden layer weights, that leads to analytical solution for the output layer weights. It makes them extremely fast for training, testing, and for generating predictions, compared to backpropagation-based neural networks, or conventional classification models. Despite their simplicity, their strength has been proven against various audio classification tasks~\cite{HanYuTashev2014}, motivating their use in this work. ELM was used for predicting audio event labels using the VTL-CNN audio embeddings as input. Each node in the ELM output layer corresponds to the probability of an utterance belonging to the corresponding class. We used Kernel-based ELMs, including radial-basis function (RBF), linear, and polynomial kernels~\cite{iosifidis2015large}.
%
\subsubsection{Support Vector Machine}
%
Support Vector Machine (SVM) is a simple classifier that constructs a set of hyperplanes in a high-dimensional space to separate data in different classes. It is based on convex optimization meant for binary classification. A one-vs-all scheme was used for performing multi-class classification out of binary SVMs. Three kernels (linear, polynomial, and RBF) were adopted for comparison studies. Notice that SVMs can be sensitive to the choice of model hyper-parameters which were optimized for each data set and choice of kernel.
%
\begin{table}[t]
\centering
\caption{Comparison of VTL-CNN audio embeddings with state-of-the-art acoustic features evaluated on ESC-50.}
\vspace{+3mm}
%
\begin{tabular}{|c|c|c|}
\hline
\textbf{Feature} & Classifier & \textbf{Mean Acc.(\%) }\\ \hline
MFCC + ZCR~\cite{piczak2015esc} & Linear SVM & 40 \\ \hline
Proposed VTL-CNN & Linear SVM & 80 \\ \hline
\end{tabular}
\label{table_results}
\vspace{-1mm}
\end{table}
%
\section{Experiments \& Results}
\label{sec:exp}
VTL-CNN audio embeddings of dimension $1024$ were extracted from audio clips after pooling the outputs of layer FC1 (Fig. \ref{fig_proposed}a-b). For LadderNet, $5$ total layers were used with three hidden layers of dimension [2048, 1024, 256]. We used Gaussian noise of $\mathcal{N}(0,0.2^2)$, with a denoising cost of [1000, 10, 0.1, 0.1, 0.1] for each layer; ReLu activations and batch normalization of size $100$ was used in all hidden layers, and the network was trained over $101$ epochs. The ELM network had a fixed number of hidden nodes corresponding to 10x the input feature dimension, with randomly initialized hidden weights. Polynomial kernels were chosen, after optimizing over all kernels and parameters. The multi-class SVM model was also optimized over different kernels and parameters, with linear kernels outperforming all other setups, evidently due to the feature space being already adequately discriminative. For ESC-50, we used the author's recommended 5-fold data split and report the average accuracy over all five folds. For CURE, fixed balanced training and test sets were defined and randomly split into 70-30\%.

Fig.~\ref{fig_results_all}a-b show results on the effect of feature normalization during training and testing for both data sets. Solid bars correspond to results obtained on the original (raw) data, and  bars with stripes correspond to the evaluation on feature-normalized data, for ESC-50 (a) and CURE (b) respectively, with y-axis showing (weighted) classification accuracy. The robustness of LadderNet and SVM models is evident against feature normalization; notice however that the SVM performance varied greatly during parameter optimization, which was not the case for LadderNet. In panel (a) ELM appears unable to learn the representation without proper normalization of the input data. This observation can be expected given the non-uniformity of the data. The effect of normalization was not as apparent for the CURE dataset (b), most likely due to the pre-treatment with perceptual correction. Further normalization of CURE at the feature level (striped bars) resulted in negligible changes in terms of accuracy for all models except ELM, which displayed a slight decrease, attributed to potential distortions introduced by mismatches between training and test set data.
Fig.~\ref{fig_results_all}c shows classification accuracy for the best proposed models, where ESC-50 set was normalized at the feature level, while CURE was not.

The proposed VLT-CNN features are compared with a popular state-of-the-art feature set comprising of spectral-based MFCC and zero crossing rate (MFCC+ZCR) features, as evaluated on one of the datasets (Table~\ref{table_results}).
In accordance to previous findings~\cite{aytar2016soundnet}, a significant  improvement is observed, validating the efficacy of transfer-learned VTL-CNN features over domain-restricted spectral-based features.
%
%
%
\section{Discussion and Conclusions}
\label{sec:discuss}
%
 An audio event classification framework was proposed consisting of CNN-based vanilla transfer learning (VTL-CNN). 
The proposed approach invoked a VTL-CNN model as feature extractor trained on  out-of-domain data, which eliminated the need for heavy retraining or introduction of additional and more complex network layers; 
Audio embeddings obtained from the pre-trained VTL-CNN were found to be more discriminative than equivalent spectral-based features; such robust features allow for efficient transfer learning and can enable the use of simpler classifiers in decision-making frameworks, when invoked with caution.
The learned feature space was combined with various decision-making models, each bringing advantages under certain conditions. When tested on an existing (ESC-50) and a newly created data set (CURE), LadderNet was proven robust to data mismatches which makes it a reliable classifier for unseen acoustic conditions. This network is also flexible enough to incorporate unlabeled data, providing an advantage for practical systems where large amounts of unlabeled data can accumulate over time. The network was further found robust to parameter tuning and data normalization, making it a great candidate for applications where high data-signature mismatch is expected.
Less complex classifiers such as Support Vector Machine (SVM) and Extreme Learning Machine (ELM), were found to be more sensitive to parameter tuning; furthermore, in the case of ELM, lack of proper data or feature normalization was proven highly unfavorable for the task of audio event classification under diminished data uniformity assumptions. Such less complex models are, however, better candidates for realistic applications constrained by time, latency and computational complexity, but extra steps for in-domain integration are necessary.
\bibliographystyle{IEEEbib}
\bibliography{MyRefs}
\noindent
\end{document}